\documentclass[sigconf]{aamas} 

\usepackage{balance} 

\setcopyright{none}
    \acmConference[ALA '25]{Proc.\@ of the Adaptive and Learning Agents Workshop (ALA 2025)}{May 19 -- 20, 2025}{Detroit, Michigan, USA, ala-workshop.github.io}{Avalos, Aydeniz, M\"uller, Mohammedalamen (eds.)}
    \copyrightyear{2025}
    \acmYear{2025}
    \acmDOI{}
    \acmPrice{}
    \acmISBN{}
\acmSubmissionID{<<submission id>>}

\title{Multicopy Reinforcement Learning Agents}

\author{Alicia P. Wolfe, Oliver Diamond, Brigitte Goeler-Slough, \\ Remi Feuerman, Magdalena Kisielinska, Victoria Manfredi}
\affiliation{
  \institution{Wesleyan University}
  \city{Middletown, CT}
  \country{United States}}
\email{pwolfe, odiamond, bgoelersloug,rfeuerman,mkisielinska, vumanfredi@wesleyan.edu}

\begin{abstract}
Inspired by the problem of optimal packet duplication in Mobile Wireless Networks, this paper examines a specific type of cooperative multi-agent problem in which an agent makes multiple identical copies of itself in order to achieve a noisy and difficult single agent task more reliably or efficiently. The agent must balance the cost of sending more copies with the improvement in speed or success rate generated by the extra copies. Due to the space of possible joint reward functions, this specific case has some differences from the more general cooperative multi-agent problem. We propose a learning algorithm for this multicopy problem which takes advantage of the structure of the value function to efficiently learn how to balance the costs and benefits of adding additional copies.

\end{abstract}

\keywords{Multiagent Reinforcement Learning, Cooperative Agents, Wireless Networks}

\newcommand{\BibTeX}{\rm B\kern-.05em{\sc i\kern-.025em b}\kern-.08em\TeX}

\usepackage{latexsym}
\usepackage{amsmath}
\usepackage{amsthm}
\usepackage{booktabs}
\usepackage{enumitem}
\usepackage{graphicx}
\usepackage{color}

\usepackage{mathtools}          
\usepackage{mathrsfs}           
\usepackage{graphicx}           
\usepackage{subcaption}         
\usepackage{url}                

\DeclareMathOperator*{\argmax}{arg\,max} 

\def\multiset#1#2{\ensuremath{\left(\kern-.3em\left(\genfrac{}{}{0pt}{}{#1}{#2}\right)\kern-.3em\right)}}

\begin{document}


\pagestyle{fancy}
\fancyhead{}


\maketitle

 \section{Introduction}

This paper introduces a new type of multiagent problem, in which a single agent duplicates itself to achieve a single goal with greater speed, quality or reliability. While this problem can be cast as a specific case of cooperative agents (for algorithms in which the number of agents can be optimized), the problem has properties that lend themselves to solutions that are tuned for this scenario.

In most multi-agent domains, agents either cooperate or compete to achieve their goals \cite{marl-book,gronauer_multi-agent_2022,wong_deep_2023}. This paper seeks to address a different type of multi-agent scenario, one in which agents duplicate themselves to combat noise and risk in the problem domain. One such domain is packet forwarding in mobile wireless networks, where duplicate packets have been shown to improve reliability and reduce delay \cite{multicopy-spyro}. Other domains include robot search and rescue, where duplicate agents may improve performance.

We assume that, in a noiseless and no risk environment, the task at hand could be completed by a single agent following an optimal policy. We further assume that if additional agents complete the task, this does not improve the outcome. However, actions are so noisy that even an agent behaving optimally has a nontrivial chance of failure. In this case, an agent may duplicate itself and make several attempts concurrently to achieve the goal. This agent must balance two factors: the cost of adding more copies vs the increase in performance due to replication. In this case, the reward function factors into a cost portion, which is summed over all agents, and an optimization portion, which is taken from only the highest performing agent. This restricts the form of the reward function of what is otherwise a cooperative multiagent setting.

We also assume that communication between agents is prohibitively expensive, except at the point where agent copies are made. When copies are made the agent(s) may jointly decide how many copies to make and what combination of actions to take: whether to take the same action with multiple copies or a range of actions to diversify the copies. After this point, we assume that noise in the domain will suffice to produce a diverse set of outcomes for different copies.

\section{Application to Wireless Networks}

In Mobile Wireless Networks, particularly sparse networks, there are cases in which not all packets make it to their destination within a reasonable time frame \cite{Jain2004:DTN}. A mobile wireless network can be viewed as a dynamic graph over a time. In a dynamic network, contiguous paths from a source to a destination may exist instantaneously or only across time. In this situation it can be useful to send more than one copy of a packet \cite{multicopy-spyro}, so that at least one copy gets to the destination in a reasonable time frame. We have designed our test bed to duplicate this scenario in some respects.

As in \cite{manfredi_learning_2024}, we treat each packet as an agent moving through an environment made up of devices. There are some properties of the Wireless Network we have constructed our gridworld to match: messages/agents must travel from a source to a goal point; only one copy of the agent must reach the goal for success and success among the copies varies due to noise in the environment.

To construct a simple gridworld in which to test our algorithms, we made some simplifying assumptions: we use tabular states and actions rather than feature based RL with machine learning; 
in the mobile network, the available actions will change as device neighbors change, whereas in our gridworld the action set in each state is fixed;
and packets affect each other indirectly through congestion in the network: actual network data will include features that reflect current network congestion levels, but no such effect is modeled in our test scenario.



\section{Reinforcement Learning}
\label{sec:rl}

Reinforcement Learning (RL) assumes an agent existing in an environment which can be represented by a Markov Decision Process (consisting of states $S$, actions $A$, and  transition probabilities between states) \cite{sutton_reinforcement_2018}. 
The agent seeks to maximize a reward signal over sequential, extended time interaction with the environment.

To do this, the agent must maximize {\em return}. For episodic tasks the return at time $t$ is defined as the sum of discounted rewards from $t$ to the end of the episode ($T$):
\begin{align}
G_t = r_t + \gamma \cdot r_{t+1} + \dots + \gamma^{T-t} r_T
\end{align}
where $r_t$ is the reward observed at time $t$, and $\gamma$ is a discount between 0 and 1 on future rewards. This return definition can be written recursively:
\begin{align}
    G_t &= r_t + \gamma \cdot G_{t+1}
\end{align}

The action value function, which is used to choose actions, can be written as a function of the expected value of the return:
\begin{align}
Q(s,a) & = E \left [ G_t \mid S_t = s, A_t = a \right ] 
\end{align}
where $S_t$ is the state at time $t$ and $A_t$ is the action at time $t$.

\section{Best Actions vs Best Multiaction}
\label{sec:variance}

Combining the cost reward for our agent copies is a straightforward sum. The optimization reward, on the other hand, is maximized over all agent copies, resulting in more interesting behavior. When splitting into multiple copies, the agent may choose a set of immediate actions for the copies to take. Always including the action with the highest individual expected value may seem best. However, a combination of actions with high expected values but low variances can lead to worse outcomes than a combination with lower expected values but higher variance.

Consider a simple Multiarmed Bandit \cite{sutton_reinforcement_2018} problem with two arms, arm Stable (S): normal distribution with $\mu = 10, \sigma = 1$ and arm Noisy (N): normal distribution with $\mu = 5, \sigma = 30$. To get the expected return of action sets like (S, N), we can generate samples from both distributions and take the max of the samples. 
Mean optimization reward returned by combinations of up to 2 arms after 10,000 samples are shown in Table \ref{tab:normals}, rounded to whole numbers.
\begin{table}[h]
\caption{Max of stable (S) and noisy (N) normal distributions.}
\begin{center}
\begin{tabular}{c|ccccc}
Action & S & N & (S,S) & (S,N) & (N,N) \\
\hline
\\
Est. Val. & 10 & 5 & 11 & 20 & 22 \\
\end{tabular}
\end{center}
\label{tab:normals}
\end{table}
Action (N, N) is the best 2-action combination, despite action N having a lower expected value. Due to higher variance, two copies of arm N are more likely to generate at least one high return.

In Table \ref{tab:constant_exp} we can see the same experiment for a bandit in which in which a mix of two different distributions is best: the two arms are a constant (C) with value 100, and exponential (E) $exp(1/70)$. 
\begin{table}[h]
\caption{Max of constant (C) and exponential (E) distributions.}\begin{center}
\begin{tabular}{c|ccccc}
Action & C & E & (C,C) & (C,E) & (E,E) \\
\hline
\\
Est. val. & 100 & 70 & 100 & 116 & 104 \\
\end{tabular}
\end{center}
\label{tab:constant_exp}
\end{table}
Here (C, E), a stable distribution combined with a noisy distribution provides the best expected maximum.




\section{Cooperative Multiagent RL}

Multiagent Reinforcement learning often starts by considering an idealized MDP that uses the cross product of the states and actions of each agent in a set of agents. In the case of cooperating agents, the agents optimize a joint reward function in this multi-agent MDP \cite{marl-book}. Work in multi-agent RL typically uses function approximation to simplify the large joint state space, or uses a view of the joint state that is only partially observed by each agent or small group of agents, as in e.g. \citet{khorasgani_k-nearest_2022,schroeder_de_witt_multi-agent_2019}. We instead begin by examining whether we need to consider the joint state and action space at all. Once copies have been created, they independently work to achieve the goal to the best of their ability. As we will see in Section \ref{sec:theory}, there is no coordination needed outside joint actions in states where copying occurs. 

\begin{figure}
    \centering
     \includegraphics[width=0.5\linewidth]{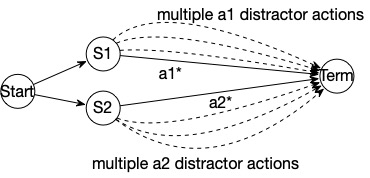}
    \caption{Shadowed Equilibrium Example.}
    \label{fig:shadowEq}
\end{figure}

{\bf Shadowed Equilibria.} Multicopy RL shares the problem of shadowed equilibria \cite{oroojlooy2023review, fulda_predicting_2007} with more general cooperative RL scenarios, however, the problem occurs differently. Consider the example MDP shown in Figure \ref{fig:shadowEq}. In a general cooperative setting we can construct a shadowed equilibrium as follows: (1) place two agents at states $S_1$ and $S_2$; (2) set the reward on the action combinations to, e.g.: $(a_1^*, a_2^*) = 10$, $(a_1, a_2) = 5$, $(a_1^*, a_2) = 1$, $(a_1, a_2^*) = 1$; and (3) have many  $a_1$ and $a_2$ distractor actions available. In algorithms where the two agents act independently, during random exploration, $a_1$ will have better estimated value than $a_1^*$ because both are more likely to be paired with one of the $a_2$ distractors. Similarly, $a_2$ will have better estimated value than $a_2^*$. Once the action combination $(a_1, a_2)$ has been learned, the agent is stuck: trying $a_1^*$ or $a_2^*$ looks even worse than during random exploration.

To construct this scenario in our optimization reward case, we cannot simply set the rewards: they are set indirectly through the maximization of 
of individual action reward distributions. However, we can indirectly construct reward distributions for the 4 actions by modifying the example from Table \ref{tab:constant_exp}. 
Higher mean distributions go to the starred actions: $a_1^* = 110$, $a_2^* = 10 + exp(1/70)$. Lower distributions go to the distractor actions: $a_1 = exp(1/70)$, $a_2 = 100$. 
\begin{table}[h]
\caption{High and low mean constant and exponential distributions.}
\begin{center}
\begin{tabular}{c|cccc}
Action &  $(a_1^*, a_2)$ & $(a_1^*, a_2^*)$ & $(a_1, a_2)$ & $(a_1, a_2^*)$ \\
\hline
\\
Est. Val. & 110 & 127  & 117 & 111 \\
\end{tabular}
\end{center}
\label{tab:high_low}
\end{table}
When we combine these distributions and take the expected maximum, constants combined with exponentials give the highest values, as shown in Table \ref{tab:high_low}. This satisfies the requirements for a shadowed equilibrium to occur when there are enough distractor actions.

\section{Baseline Multiagent Algorithm}
\label{sec:multiagent}
As a baseline to compare to our algorithm (see Section \ref{sec:theory}), we implemented a multiagent solution with a basic commuication protocol and reward function, referred to as the Joint Action Multiagent algorithm. As a standard multiagent problem, the task has the following properties: (1) Agents may only communicate when they split into multiple copies. 
(2) Only the best agent to achieve the goal gets the reward for doing so.
(3) All agents accumulate cost rewards. 
This means that while the overall task is cooperative, individual agents compete to best achieve the goal.

While this reward function accurately reflects the task, the reward received by an agent depends on the other agents: if agent copy 2 does better at the task, then agent copy 1 receives no reward for success. Therefore, the number of copies made and the actions they took in the splitting state affect the value function for each agent. Each agent must remember the joint action chosen at the start state, or operate in a POMDP. If $n$ agents choose the action set $\{a_0, a_1, \dots, a_n\}$ in the start state, then at every state $s$ in the episode we add this information to the state, forming the augmented state $(s, (a_1, a_2, \dots a_{n}))$. 
Once the state space is augmented, each agent runs q-learning with a shared value function between all agents. When calculating the value of a candidate set of actions $(a_1, a_2, \dots a_n)$, the values for each individual action in the context of the joint action are summed. The action combination with the highest value is chosen. From that point forward, the agents act independently, even if they are on similar paths.



\section{Multicopy Actions}
\label{sec:theory}
Ideally, we'd like each agent to attempt the task to the best of its ability, despite the presence of other copies. We would therefore like to consistently reward each copy for learning to achieve the goal. However, when calculating the value of a joint action, we should only include the optimization reward received by the best copy. Our algorithm does so, and allows us to treat the domain as an MDP without remembering the joint action chosen.

We divide the reward function into two portions: cost, which reflects the additive costs of running more copies; and optimization, which reflects the improvements in success 
with additional agent copies. 
The algorithm should find the best trade-off between the two.

At certain states the agent may make up to $n$ copies of itself. In state $s$, considering multi-actions that include up to $n$ actions, the set of multi-action combinations available is the set of multisets of cardinality less than $n$ that can be formed from the set of actions:
\begin{align}
    \mathcal M (s) = \bigcup_{k = 1}^n \multiset{\mathcal A(s)}{k} 
\end{align}
where $\mathcal{A}(s)$ is the set of actions available at $s$. In some experiments we remove multi-actions with duplicate actions from consideration, while in others we include them.

The agent starts with a single copy, agent $0$. If a copy $i$ chooses multiaction $m_t$ at time $t$, each agent copy produced is assigned a unique id. Define $ID_t$ to be the set of unique agent ids at time $t$, and $ID_t(i)$ to the the set of agent ids that are created from agent $i$ at time $t$. 
$m_t$ consists of a set of actions $a_{ij}$ indexed by the original agent $i$ and the agent $j$ which results from the action: $m_t = \{ a_{ij} \mid j \in ID_t(i)\}$. 


{\bf Return Definition for Multiple Actions.}
Our goal is to avoid using multi-sets of next states to tabulate the value function, and to use multi-sets of actions only in states where duplication of the agent is considered. We consider problems where the total sample return $G(t, i)$ for agent $i$ at time $t$ can be factored into two components, the cost return $G_c(t, i)$ and the optimization return $G_o(t, i)$:
\begin{align}
\label{math:Gt}
    G(t, i) = G_c(t, i) +  G_o(t, i)
\end{align}

Costs are summed across all copies of the agent. Let $R_{c}(t, i)$ be the cost reward at time $t$ for agent $i$, and $G_{c}(t, i)$ be the cost return for agent $i$ at time $t$. The return for agent $i$ at time $t$ can be written in terms of the return for the set $ID_t(i)$ of agent copies created:
\begin{align}
    G_{c}(t, i) 
                &= \sum_{j \in ID_t(i)} \left [  R_{c}(t, j) + \gamma  G_c(t+1, j) \right].
\end{align}

In episodic problems, different agent copies may finish at different times.
$G_c$ for each agent ends at $T(i)$, the last timestep in which that agent is active:
\begin{align}
    G_{c}(T(i), i) &= \sum_{j \in ID_{T(i)}(i)} R(T(i), j).
\end{align}

Optimization rewards are only counted on the ``best" path found. Therefore at each timestep they are maximized over the returns from the next timestep. Let $R_{o}(t, i)$ be the optimization reward at time $t$ for agent $i$, and $G_{o}(t, i)$ be the corresponding optimization return. When calculating overall optimization reward for the set of agent copies, we only count the best agent's return:
\begin{align}
    G_{o}(t, i) 
    &= \max_{j \in ID_t(i)} \left[ R_{o}(t, j) + \gamma  G_o(t+1, j) \right].
\end{align}

$G_o$ for agent $i$ ends at $T(i)$, the last timestep in which $i$ is active:
\begin{align}
    G_{o}(T(i), i) &= \max_{j \in ID_{T(i)}(i)} R(T(i), j).
\end{align}

{\bf Policy Evaluation for Multiple Actions.}
The value of multiaction $m$ in state $s$ is its expected value, at any time steps where $s, m$ is experienced:
\begin{align}
    Q(s, m) =~& E \left[ G(t, i) \mid S_{t,i} = s, M_{t,i} = m \right]\\
    =~& E \left [G_c(t, i) + G_o(t, i) \mid S_{t,i} = s, M_{t,i} = m \right] 
\end{align}
where $S_{t, i}$ is the state at time $t$ for agent $i$, and $M_{t, i}$ is the multi-action at time $t$ for agent $i$.

This expectation factors into the sum of the cost expectation and the optimization expectation:
\begin{align}
    Q(s, m) =~& E \left [ G_c(t, i) + G_o(t, i) \mid S_{t,i} = s, M_{t,i} = m \right ] \\
    =~& E \left [ G_c(t, i) \mid S_{t,i} = s, M_{t,i} = m \right ] + \\ 
    ~& 
    E \left [ G_o(t, i) \mid S_{t,i} = s, M_{t,i} = m \right ]
\end{align}
where $S_{t, i}$ is the state of agent $i$ at time $t$, and $M_{t, i}$ is its multiaction.

This factors into a value function for cost, $Q_c$, and a value function for optimization, $Q_o$: 
\begin{align}
    Q(s, m) =~& Q_c(s, m) + Q_o(s, m) 
\end{align}
The cost q-function factors further, into a sum of q-values for individual actions $a \in m$:
\begin{align}
    Q_c(s, m) =~& E \left [ G_c(t, i) \mid S_{t,i} = s, M_{t,i} = m \right ]\\
    =~&  E \left[  \sum_{j \in ID_t(i)} G_c(t, j) \mid S_{t, i} = s,  \in M_{t, i} = m \right ] 
\end{align}
If $A_{t, j}$ is defined as the individual action that produces agent copy $j$ at time $t$, then this becomes:
\begin{align}
    =~&  \sum_{a \in m}E \left[   G_c(t, j) \mid S_{t, i} = s,  A_{t, j} = a \right ] \\
    =~& \sum_{a \in m} Q_c(s, a)
\end{align}
Therefore, for $Q_c$ we can use any learning technique, including bootstrapping methods like Q-learning \cite{watkins1992, sutton_reinforcement_2018} or SARSA \cite{sarsa_rummery1994, sutton_reinforcement_2018} to learn the Q-function for individual actions, and sum for multiactions. 

$Q_o$ cannot be simplified in the same way, as the expected value and maximization do not commute in general:
\begin{align}
    Q_o(s, m) = &~E \left[  \max_{j \in ID_t(i)}G_o(t, j) \mid S_{t, i} = s,  M_{t, j} = m \right] 
\end{align}
Here we must learn $Q_o$ for each combination of actions we consider. However, within the maximization step each $G_o$ may be calculated individually, allowing us to avoid a value function defined over the cross-product of states. If we are estimating $G_o$ over multiple times steps this does limit us to techniques which can be used without bootstrapping like Markov Chain Monte Carlo \cite{sutton_reinforcement_2018}.

{\bf Policy Improvement.}
When calculating the best action, the agent should optimize the total value:
\begin{align}
\pi(s) = &\argmax_m E \left [G_c(t, i) + G_o(t, i) \mid  S_{t, i} = s,  M_{t, i} = m \right ].
\end{align}
This function cannot be maximized separately over cost and optimization values, therefore action selection must be done using the joint value function and its policy.


{\bf Multicopy agent.} The multicopy agent combines a cost agent (\(Q_c\)), and an optimization agent (\(Q_o\)), and sums their values to calculate the joint multicopy policy \(\pi\). 

{\bf Cost and Optimization Agents.}
The cost agent  estimates \(Q_c\) for each state $s$ and individual action $a$ using Q-learning. To calculate the value of a multiaction $m$, these are summed. The optimization agent estimates \(Q_o\) for each state $s$ and multiaction $m$, using Every Visit Markov Chain Monte Carlo \cite{sutton_reinforcement_2018}.

\begin{figure}
    \centering
    \hfill
    \begin{subfigure}{0.45\textwidth}
     \includegraphics[width=0.9\textwidth]{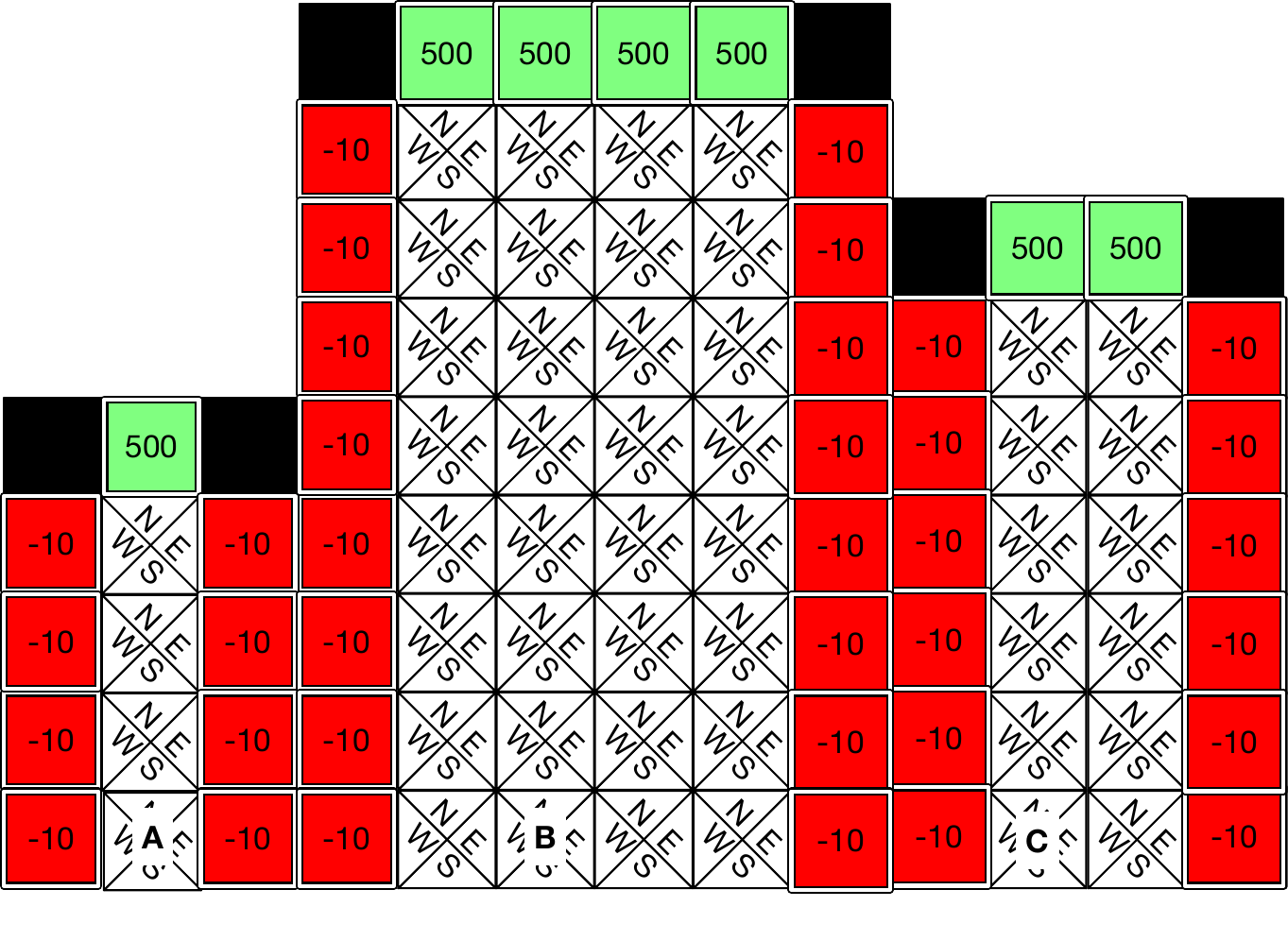}
    \caption{{\bf Three Bridges:} quick, risky bridge (A); slow, safe bridge (B); medium bridge (C).}
    \label{fig:threeBridges}
    \end{subfigure}
    \hfill
    \begin{subfigure}{0.45\textwidth}
        \includegraphics[width=0.9\textwidth]{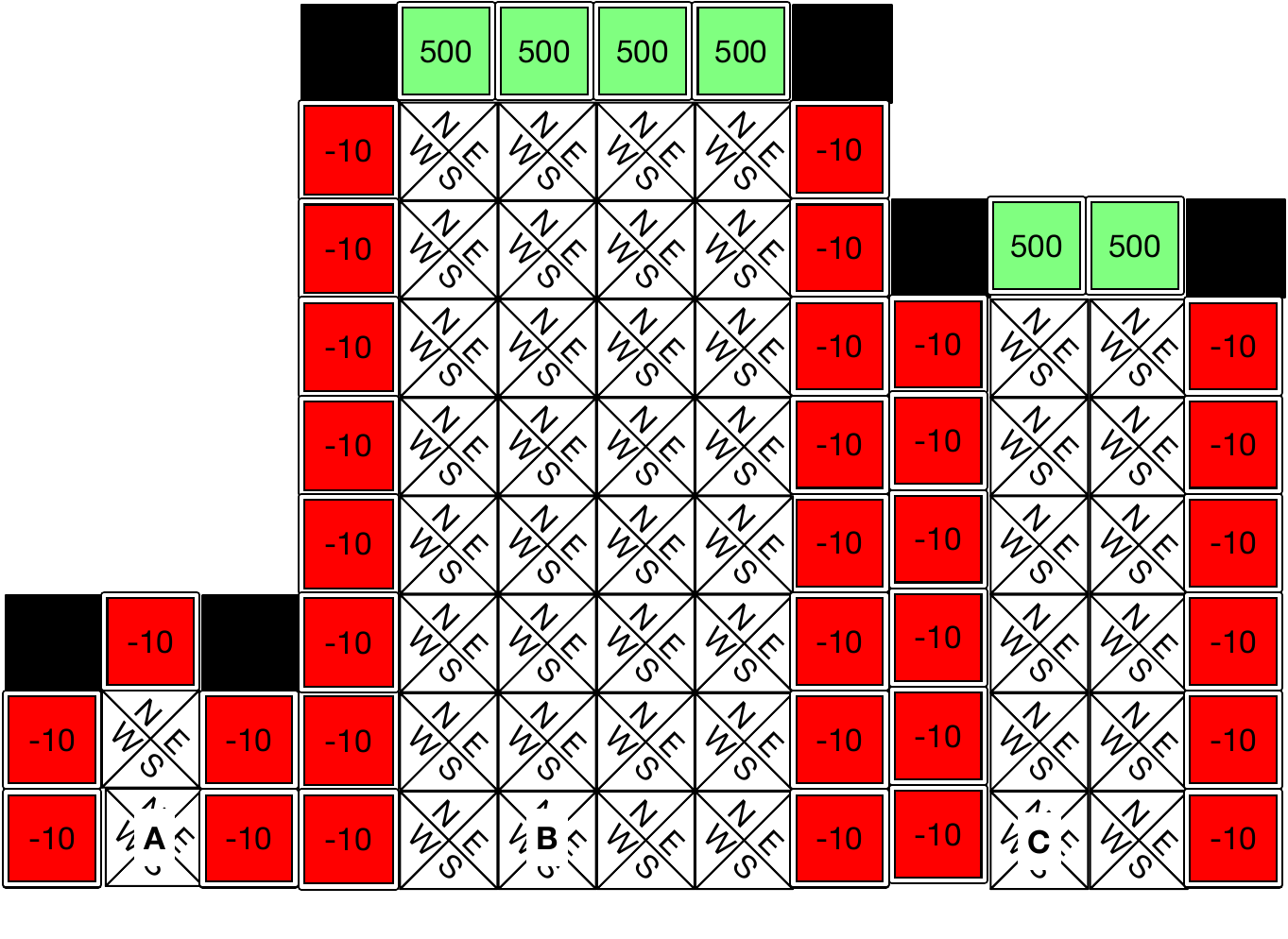}
    \caption{{\bf Bridges Gridworld} with one broken bridge (A). The broken bridge is randomly chosen per episode.}
    \label{fig:BBGridworld}
    \end{subfigure}
\caption{Three-bridge gridworlds. Terminal states are marked with double squares. All other states have 4 actions: North, South, East, West. Postive terminal state rewards are optimization rewards, negative are costs. White squares have a per-step cost that is set for each experiment.}
\end{figure}

\section{Gridworld Domains}

The testing domains were chosen to examine the ability of the Multicopy agent to adapt the number of agent copies and the actions those copies take to improve performance under various levels of noise and cost.

{\bf Gridworld Environment.} To model tasks with varying risk, delay and cost in a simple domain, we use a set of gridworld bridges of varying length and width, shown in Figure \ref{fig:threeBridges}. This is an expanded version of a single bridge gridworld from the Pacman Projects \cite{pacman_2010}. Falling off a bridge represents task failure, and successfully making it across any bridge represents task success. 

The three bridges present different trade-offs between speed and risk. Bridge (A) is short but narrow, risky when noise is high. Bridge (B) is long and wide: safer, but slower. Bridge (C) is a compromise.

{\bf Start state.} In the start state (not shown), the agent may choose any single bridge or a combination of bridges, up to a maximum number of actions (set to 1, 2 or 3 per experiment). 
The actions in the start state have no noise, transitioning to the states labeled "A", "B" or "C" in Figure \ref{fig:threeBridges}. An agent copy cannot transition back to the start state once it has chosen a bridge.  

{\bf Noise.} The gridworld includes an adjustable noise parameter \cite{pacman_2010}. With noise setting $\beta$, directional actions (N, S, E, W) on the bridges  go in the intended direction with probability $1-\beta$, and go in the two orthogonal directions with probability $\beta/2$ each. 

\begin{figure*}
    \centering
    \begin{subfigure}{0.3\textwidth}
         \centering
    \includegraphics[width=\textwidth]{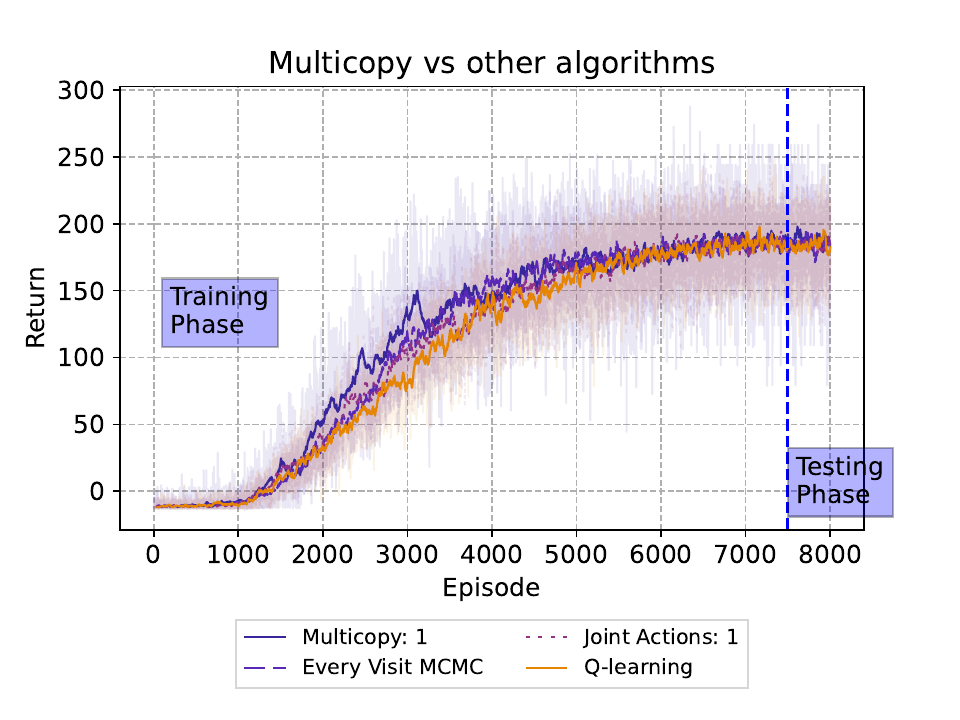}
         \caption{Basic learning: Noise 0.1}
         \label{fig:basic_noise0}
     \end{subfigure}
     \hfill
    \begin{subfigure}{0.3\textwidth}
         \centering
         \includegraphics[width=\textwidth]{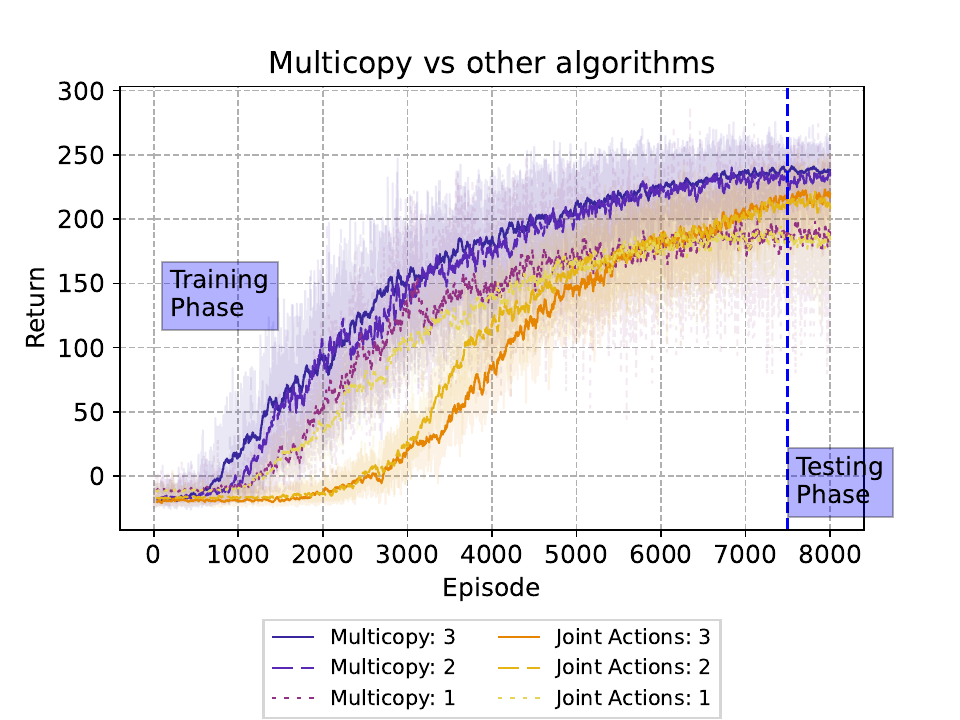}
         \caption{Multiple Actions: Noise 0.1}
         \label{fig:value_noise0}
         \vfill
     \end{subfigure}
    \hfill
         \begin{subfigure}{0.3\textwidth}
         \centering
         \includegraphics[width=\textwidth]{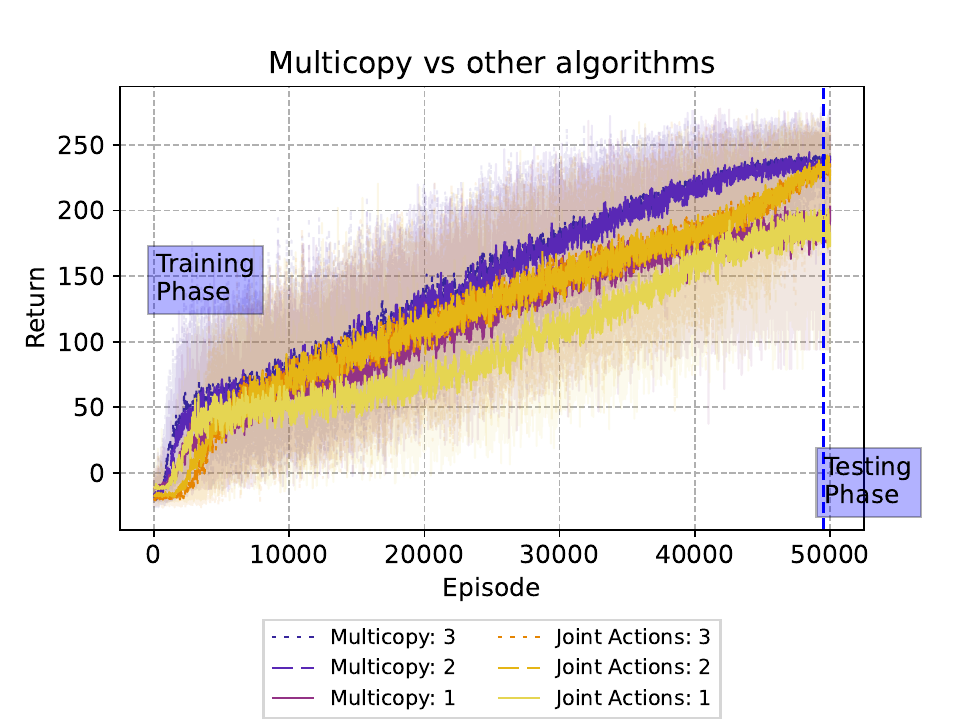}
\caption{Long term: Noise 0.1}
         \label{fig:action_noise0}
     \end{subfigure}
\begin{subfigure}{0.3\textwidth}
         \centering
    \includegraphics[width=\textwidth]{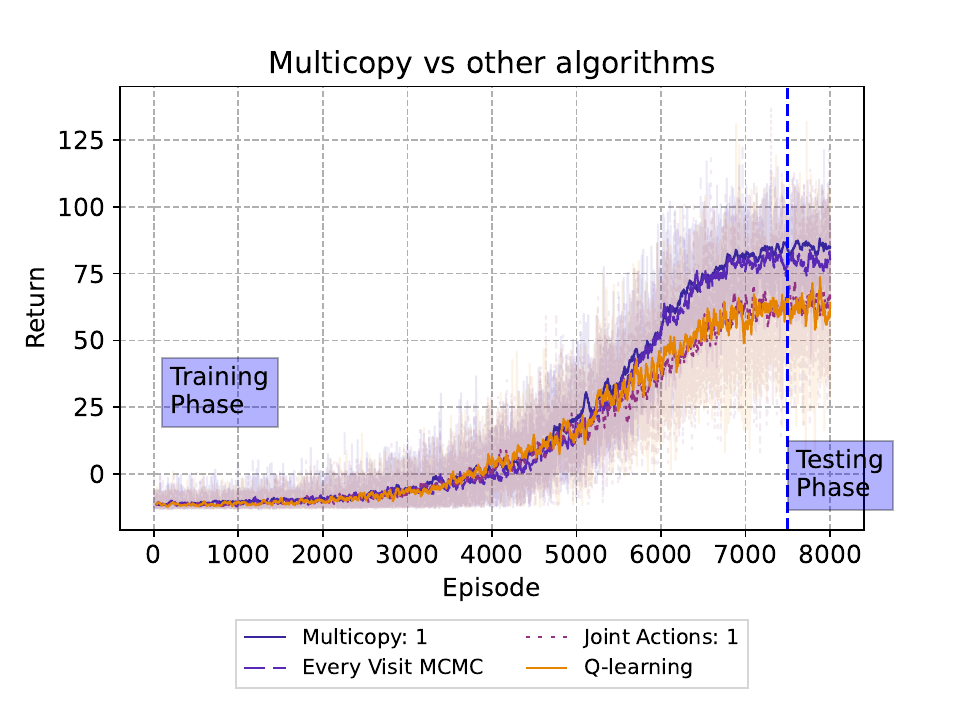}
         \caption{Basic learning: Noise 0.3}
         \label{fig:basic_noise3}
     \end{subfigure}
     \hfill
    \begin{subfigure}{0.3\textwidth}
         \centering
         \includegraphics[width=\textwidth]{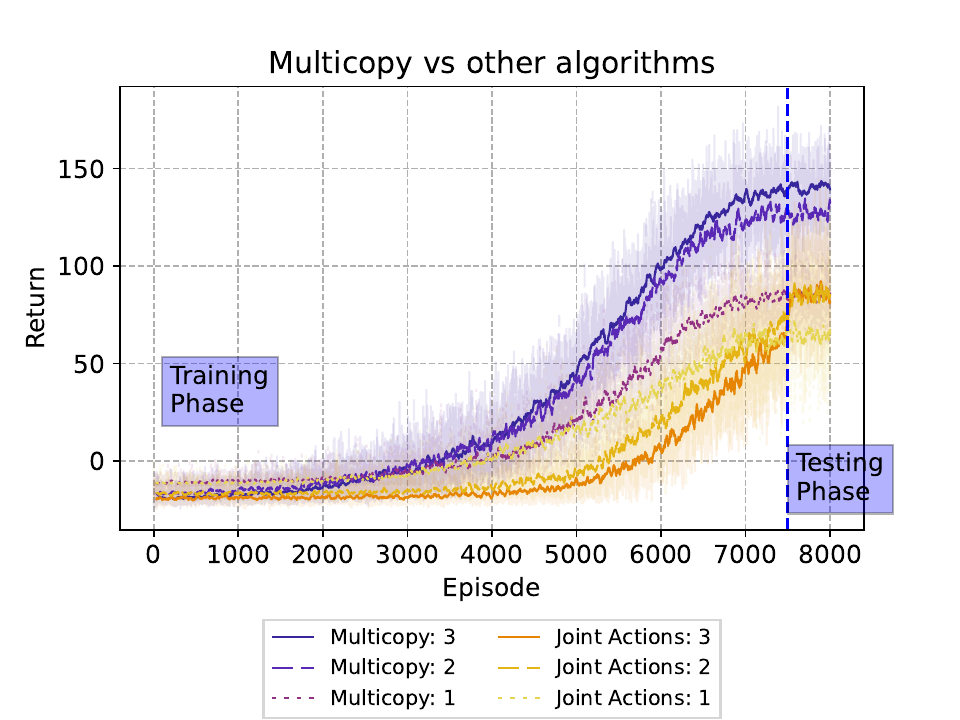}
         \caption{Multiple Actions: Noise 0.3}
         \label{fig:value_noise3}
         \vfill
     \end{subfigure}
    \hfill
         \begin{subfigure}{0.3\textwidth}
         \centering
         \includegraphics[width=\textwidth]{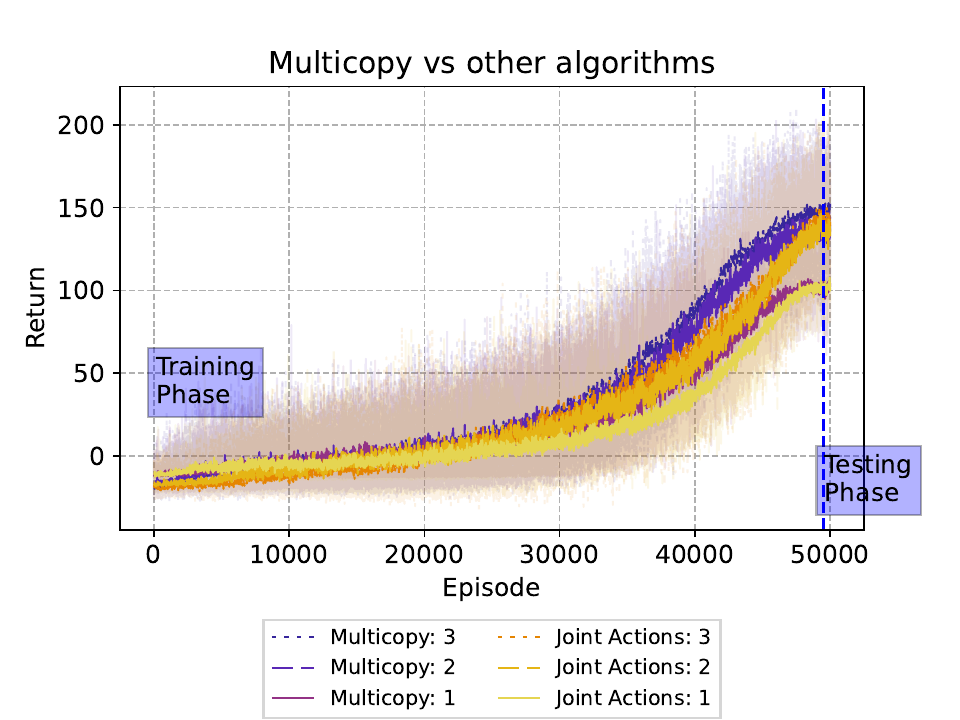}
\caption{Long term: Noise 0.3}
         \label{fig:action_noise3}
     \end{subfigure}
     \caption{{\bf Three Bridges:} Learning results for basic non-multicopy vs multicopy algorithms on the 3 bridges gridworld. Cost per step is -2 in all plots. Shaded lines are averaged over 50 trial runs, dark lines are rolling 30-episode averages of those values.
     }
    \label{fig:basic_compare}
\end{figure*}

{\bf Optimization and Cost.} The optimization reward, shown as positive reward in Figure \ref{fig:threeBridges}, rewards successful crossings. The cost reward consists of a small negative reward on each step in the white bridge states, and a larger negative reward (shown) for falling off the bridge. The cost in bridge states varies per experiment. 


{\bf Broken Bridges.} 
In the version of the three bridge gridworld shown in Figure \ref{fig:BBGridworld}, at the start of each episode one randomly chosen bridge is broken midway along (Bridge A in the diagram). All agents crossing the broken bridge during the episode will fail.


\section{Experiments}

The experiments in this section aim to determine if the algorithm can find the right number of agent copies and combination of actions. First, we compare our algorithm from Section \ref{sec:theory} to the basic algorithm with joint actions from Section \ref{sec:multiagent} in the Three Bridges Gridworld, with various maximum action settings. Then we compare the number and type of actions chosen for various cost and noise settings, to confirm that the algorithm is able to adjust the multiactions chosen according to the domain. Next we examine the difference in policy when we allow duplicate actions (more than one agent on the same bridge), and finally, some more nuanced experiments with duplicate actions.

{\bf Exploration.}
On-policy techniques like MCMC can be overly cautious when noise is high. 
We therefore use a boltzmann distribution with action advantages and linearly decreasing temperature over time, starting at 100 and ending at 1, for exploration. Temperature decreases at the end of each episode.

{\bf Algorithm Parameters.}
The learning rates for MCMC and Q-learning typically differ, with lower learning rates for MCMC. In all algorithms we use 0.05 for MCMC and 0.2 for Q-learning, linearly decreasing to 0 through the training episodes.
We used a discount of 0.9 for all experiments. 

{\bf Training and Testing.} Each agent was trained for 7500 episodes and tested for 500 episodes, except where noted. During the testing phase, exploration and learning are turned off completely and only the optimal multiaction is chosen.

\subsection{Learning Curves} 
Figure \ref{fig:basic_compare} contains learning curves for noise levels of 0.1 and 0.3, with bridge step-costs of -2. In Figure \ref{fig:basic_compare}(a) and (d), we compare our multiagent baseline algorithm (Joint Actions) and our Multicopy algorithm with MCMC and Q-learning. For this comparison, the maximum number of actions is set to 1. Note that the joint action algorithm is similar to Q-learning (on-policy) while the multicopy algorithm is closer to MCMC (off-policy).

In Figure \ref{fig:basic_compare}(b -c) and (e-f), we compare Multicopy and Joint Action algorithms with various maximum action settings. Figures (b) and (e) use the 7500 episode training time that we will use throughout the remainder of the paper, while Figures (c) and (f) use a longer training time of 50,000 episodes. The Multicopy algorithm does almost as well in 7500 episodes as it does in 50,000, even with high noise levels. The Joint Action algorithm, however, due to its fractured state space, does not learn as well when training time is short, particularly in (e), where noise is high.

\begin{figure*}
     \centering
        \begin{subfigure}{0.9\textwidth}
         \centering
         \includegraphics[width=\textwidth]{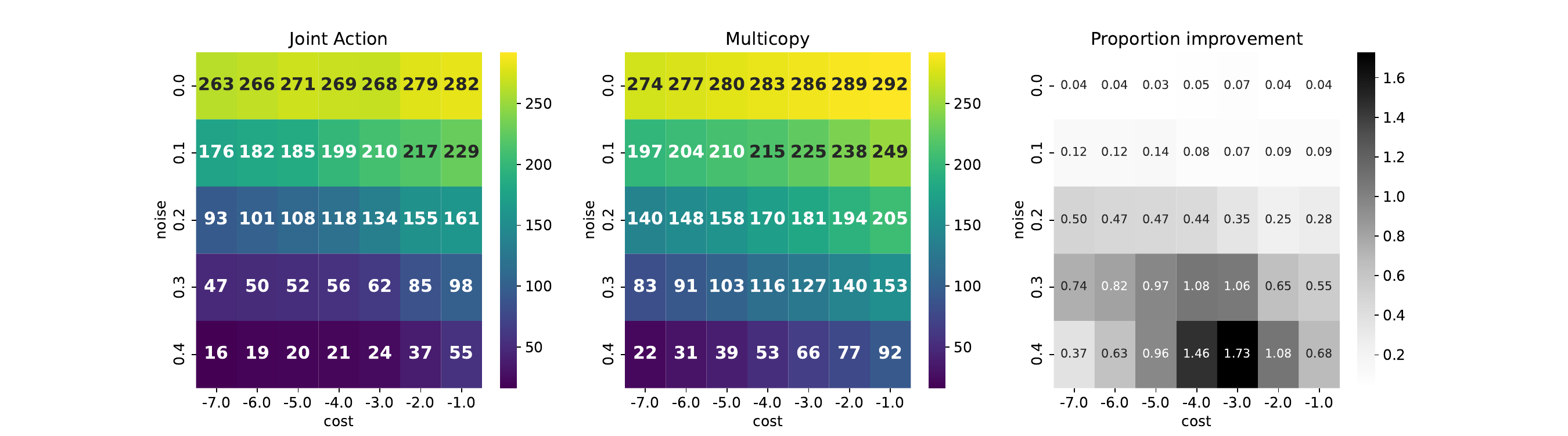}
         \caption{{\bf Return.} Per Episode return during testing phase for various noise and cost settings. Final graph shows proportion improvement from the Joint Action return to the Multicopy return.}
         \label{fig:basic_costnoise_v}
     \end{subfigure}
     \begin{subfigure}{\textwidth}
         \centering
    \includegraphics[width=0.67\textwidth]{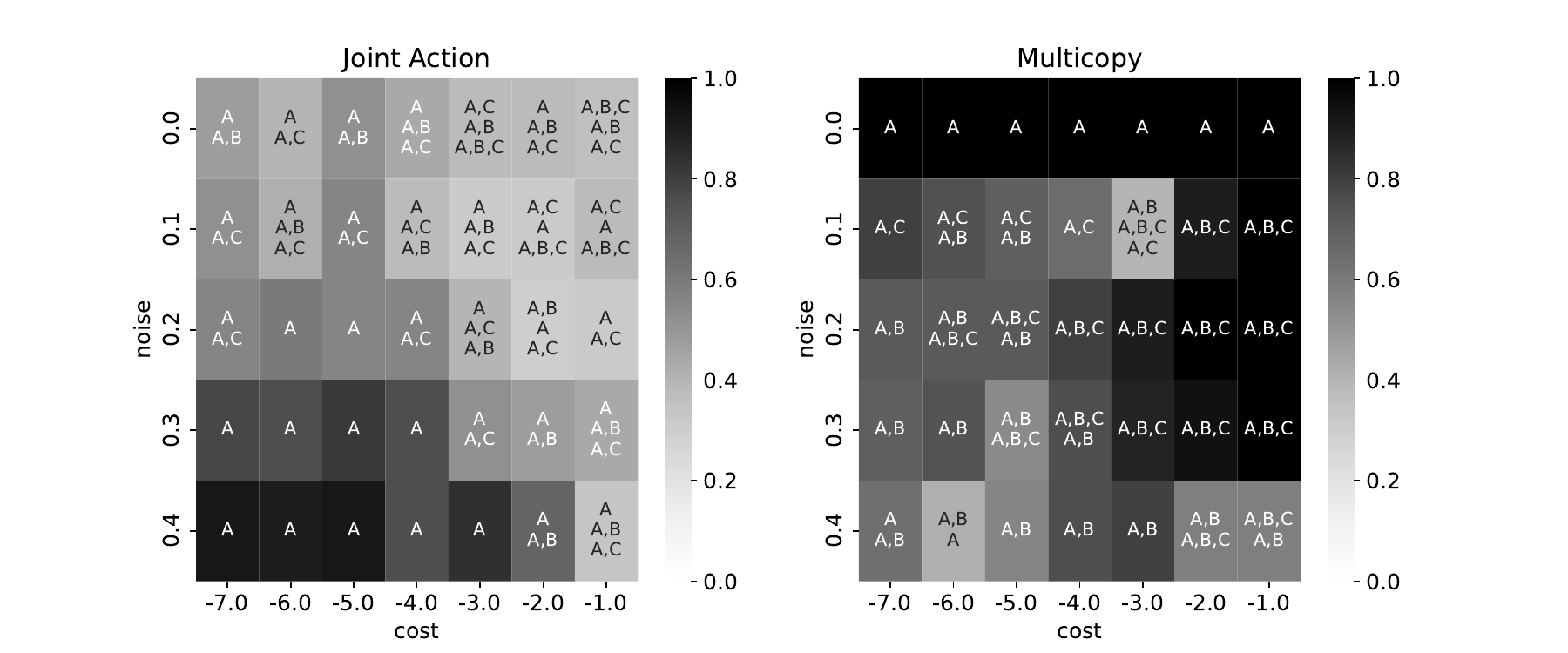}
         \caption{{\bf Policy.} Best actions in the start state for various noise and cost settings. Top action listed is chosen in most trials, shading shows proportion of trials in which this action is best. All actions best in $> 20\%$ of trials are shown in decreasing order. }
         \label{fig:basic_costnoise_pi}
     \end{subfigure}

    \caption{{\bf Varying Noise and Cost.} Return and policy during testing for various cost and noise settings, showing the improvement in return from the Joint Action algorithm to Multicopy (a). Actions shown in (b) are best actions. 50 trial runs.}
    \label{fig:basic_costnoise}
\end{figure*}

\subsection{Varying noise and costs}


This section examines the Multicopy algorithm in a variety of cost/noise combinations, both to compare to the Joint Actions algorithm, and to examine what conditions lead to choosing different combinations of actions. In these initial experiments, duplicate actions are not allowed: only one agent may be sent on each of the three bridges. These experiments include limited training time of 7500 episodes.

As shown in Figure \ref{fig:basic_costnoise_v}, the Multicopy algorithm generally performs better than the Joint Actions algorithm in these experiments. The highest proportional improvement is seen when noise is high (0.4) and costs are moderate (-4), and the lowest improvement when noise is 0.0. Figure \ref{fig:basic_costnoise_pi} details the policy chosen in the start state by each algorithm. The Joint Action algorithm is much more likely to choose a simple policy with 1 or 2 actions, perhaps because at this stage it has a better policy for getting across the bridges under those conditions.

\begin{figure*}
    \centering
    \centering
        \begin{subfigure}{0.9\textwidth}
         \centering
         \includegraphics[width=\textwidth]{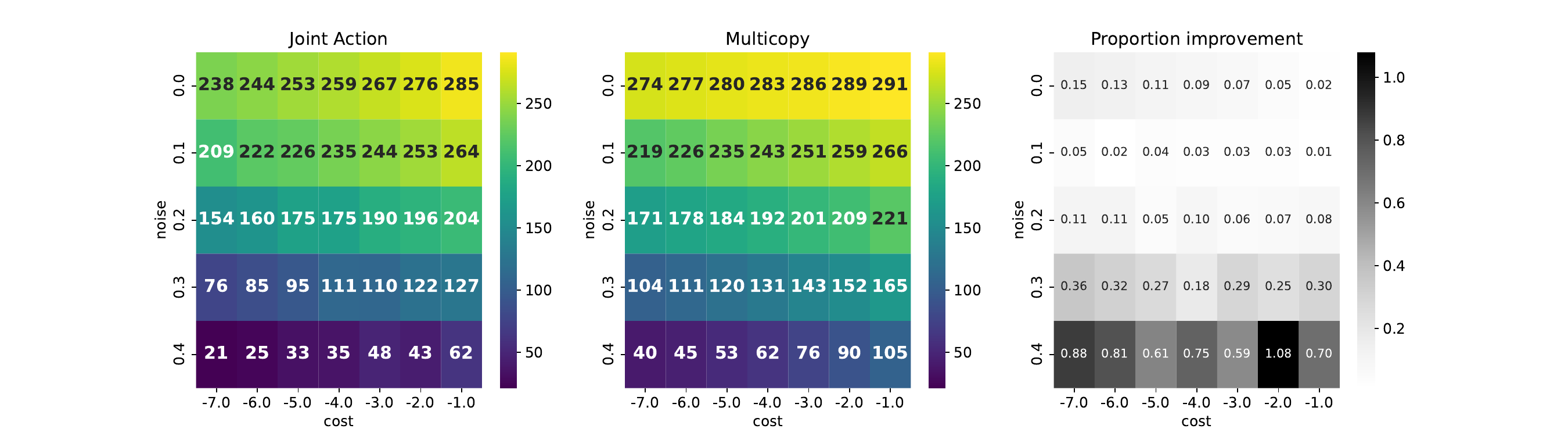}
         \caption{{\bf Return.} Per Episode return during testing phase for various noise and cost settings.}
         \label{fig:return_dups}
     \end{subfigure}
    \begin{subfigure}{\textwidth}
         \centering
    \includegraphics[width=0.67\textwidth]{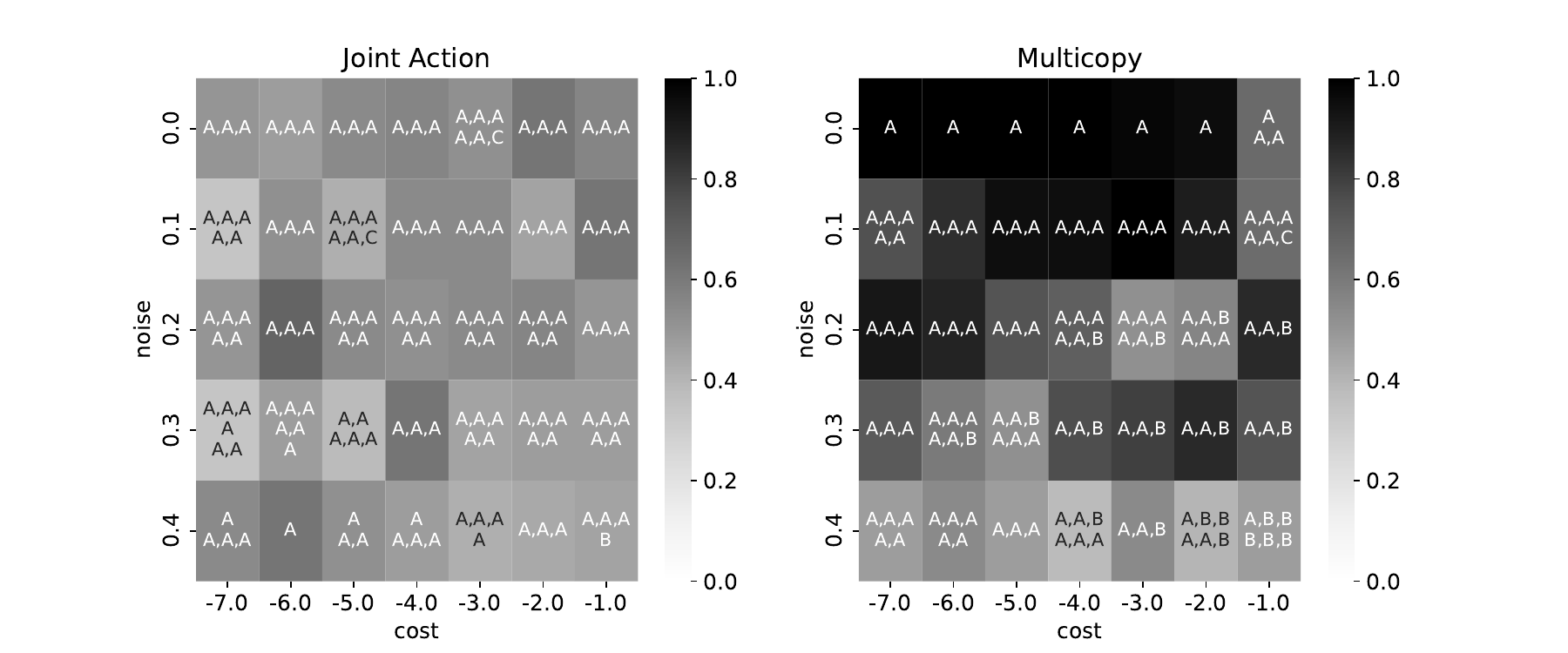}
         \caption{{\bf Policy.} Best actions in the start state for various noise and cost settings. Top action listed is chosen in most trials, heatmap value is proportion of trials in which this action is best. All actions best in $> 20\%$ of trials are shown in decreasing order. }
         \label{fig:policy_dups}
     \end{subfigure}
    \caption{{\bf Identical Duplicate Actions.} Return and policy during testing for various cost and noise settings when identical duplicate actions are allowed. 50 trial runs.}
    \label{fig:duplicates}
\end{figure*}

The Multicopy algorithm has a more complex pattern of action choices. Generally one would expect that as cost magnitude increases, the agent should make do with fewer copies. Noise should have roughly the opposite effect: as noise increases, the agent should send more copies. This rule of thumb largely applies for the upper and right rows of the graphs in Figure \ref{fig:basic_costnoise_pi}. When there is no noise, one action is consistently sufficient (first row). The area where three actions are best forms a wedge from the right side of the table, where costs are low, with a point at about -4, 0.2, where (A, B) and (A, B, C) have roughly the same value.

The number of actions goes back down to one when noise and costs are both high (the extreme lower left region). In this case the agent cannot reliably achieve the task even using all three bridges, and largely returns to using a single copy to keep costs down. Between the regions where 3 actions are best, and the regions where 1 action is best, there are bands in which 2 actions is the optimal policy.


In the initial experiments, the agent may only send a single agent on each of the three bridges, which limits the advantage to be gained from using all three agent copies. In this section we look at whether sending more than one agent on the same bridge can improve performance.

As in the previous section, we compare the performance of the Joint Action and Multicopy algorithms when a maximum of 3 actions are allowed in the start state.

\subsection{Duplicate Action Choices}

{\bf Identical Actions.}
When we remove the requirement that there be only one agent per bridge in Figure \ref{fig:duplicates}, we see that the agent tends to choose to send multiple agents over bridge A, the shortest bridge (Figure \ref{fig:policy_dups} Multicopy). This bridge is  dangerous due to its narrowness, and there is no correlation between the success or failure between agent copies crossing the bridge. This means that making multiple identical attempts at the same bridge can improve performance even if the agents do no coordination or cooperation on the bridge.

For our Multicopy algorithm, Figure \ref{fig:policy_dups} shows largely cases in which the agent chooses the maximum 3 copies, though there are still a few cases where the agent chooses fewer copies to reduce costs. Multiple copies also do not help when there is no noise in the environment (the first row of all heatmaps in Figure \ref{fig:duplicates}). While the Joint Action algorithm is closer in return to the Multicopy algorithm in these experiments (Figure \ref{fig:return_dups}), it does make the odd choice to send three agent copies even when there is no noise. Generally the "best" action is not consistent across trial runs for this algorithm, however (see Figure \ref{fig:policy_dups}), indicating that it has not fully converged.


\begin{figure*}
    \centering
        \begin{subfigure}{0.9\textwidth}
         \centering
         \includegraphics[width=\textwidth]{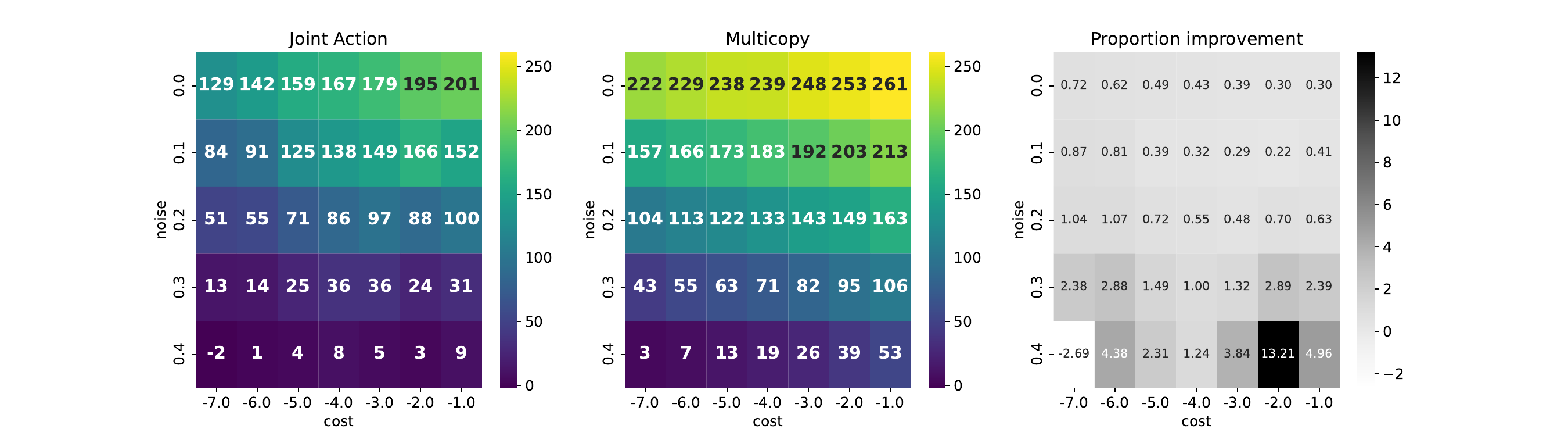}
         \caption{{\bf Return.} Per Episode return during testing phase for various noise and cost settings.}
         \label{fig:broken_costnoise_v}
     \end{subfigure}
    \begin{subfigure}{\textwidth}
         \centering
    \includegraphics[width=0.67\textwidth]{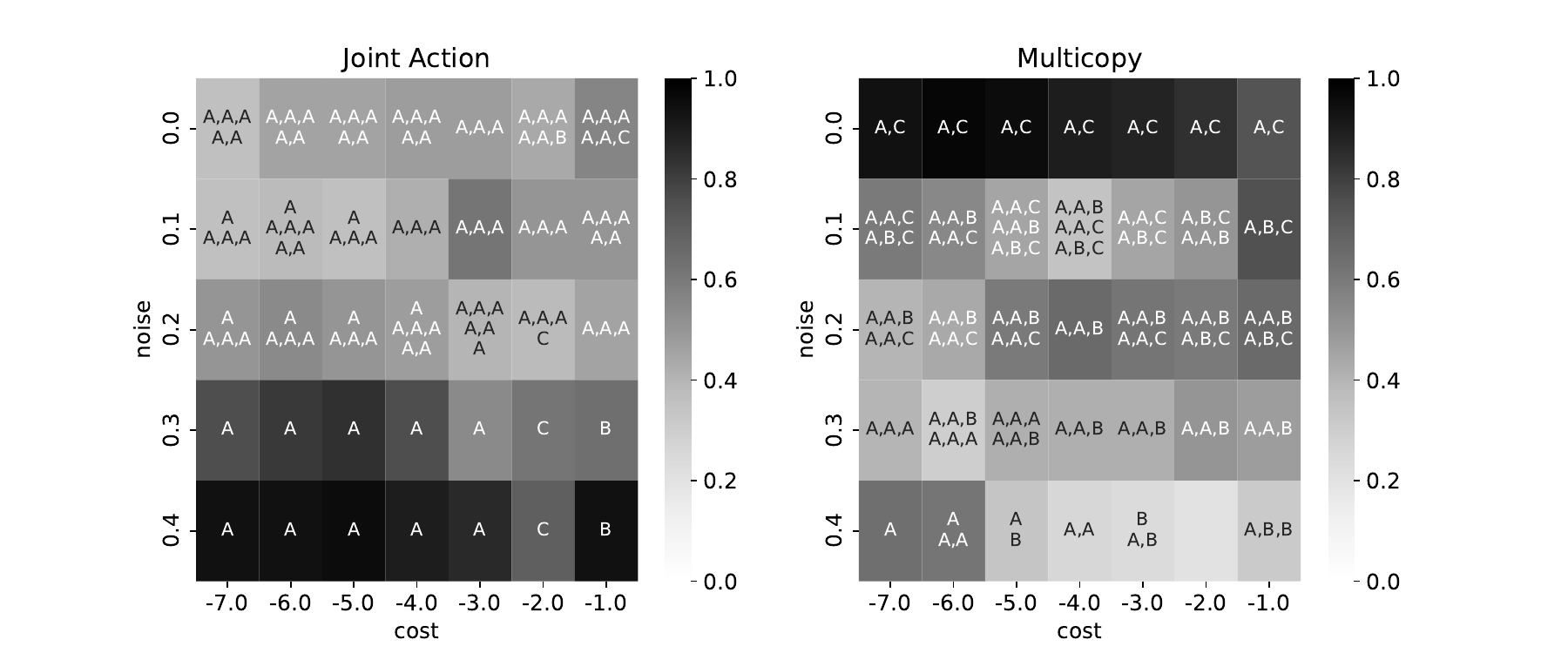}
         \caption{{\bf Policy.} Best actions in the start state for various noise and cost settings. Top action listed is chosen in most trials, heatmap value is proportion of trials in which this action is best. All actions best in $> 20\%$ of trials are shown in decreasing order. }
         \label{fig:broken_costnoise_pi}
     \end{subfigure}
    \caption{{\bf Broken Bridges Experiments.} Return and policy during testing for various cost and noise settings when identical duplicate actions are allowed and a random bridge is broken in each trial. 20 trial runs.}
    \label{fig:broken}
\end{figure*}

\subsection{Correlated Dangers}

Thus far, we have examined a case in which the agent is allowed to send only one copy per bridge 
and a case in which the agent can and does send multiple copies on the same bridge. Next we examine a case in which despite having the option to send multiple copies on the same path, the rational behavior is to spread copies over different paths. One such scenario occurs when the noise or dangers on a path are correlated for all agents on the same path.

We model this using the Broken Bridges gridworld in Figure \ref{fig:BBGridworld}. Here, each time the task is attempted, the agent finds that a random bridge is broken, and all agents on that same bridge fail at the task. This might correspond in more realistic domains to faults in parts of the wireless network, or the possibility of tunnel collapse in a multi-robot region for a robot navigating a collapsed building.

This appears to be a particularly difficult task for the Joint Actions agent to learn in the time allotted: in Figure \ref{fig:broken_costnoise_v} we can see that the Multicopy algorithm can do up to 12 times better in cases where noise is high. In Figure \ref{fig:broken_costnoise_pi} we can see that this basic algorithm often still chooses only repetitions of bridge A, which is the shortest bridge and often the first that it learns to cross.

Since sending multiple agents on the same bridge can lead to both agents being lost, the Multicopy agent prefers to spread the copies over at least two different bridges to ameliorate risk, as shown in Figure \ref{fig:broken_costnoise_pi}.

\section{Future Work}


We will continue  developing these algorithms in combination with modern function approximation methods such as Neural Networks \cite{deeprl,wong_deep_2023} in order to apply them to Mobile Wireless Networks \cite{tie2011r3,multicopy-spyro,manfredi_learning_2024}.

In this paper, and in the Mobile Wireless Networks applications that inspired it, the agent seeks to optimize the amount of time required to complete a task. However, other types of optimization functions are possible within the same mathematical framework: building the best object, or getting the farthest in an exploration or rescue task. 

In this work, we assume that communication between agents is expensive or unavailable much of the time, aside from the point at which the agent decides to duplicate itself. We therefore use a model where communication is limited to only the duplication step. 
This was instrumental in producing the simplified value function we use here. 
However, we will look for types of communication that are possible while maintaining this value function model. Once there is some level of interaction between agents, non-deterministic policies may be required \cite{vieillard_munchausen_2020,haarnoja_reinforcement_2017}.

\balance

Once we move to larger problems and approximate solutions, distributional RL \cite{bellemare_distributional_2023} may prove useful, given a maximization operator for their particle based value representation.

In this paper we primarily looked at different cost/optimization reward ratios in order to show that different multiactions are chosen at different settings. However, even in the single agent case, multiobjective RL \cite{mossalam_multi-objective_2016,moffaert_multi-objective_2014,nguyen_multi-objective_2020,hayes_brief_2023} would be a useful tool to find a range of policies for different weights. Lexicographic ordered RL \cite{tercan_thresholded_2022,hayes_dynamic_nodate}could be particularly useful if it allows us to find the best delay given a particular target for throughput, for example.



\section{Conclusion}

This paper makes several contributions. First, we have shown that for the specific multiagent problem we examine here, considering the joint state space of all agent copies is not necessary, and joint actions are necessary only when duplicating agents.

Even using a simple very gridworld, the algorithm found a complex pattern of optimization that is easier to construct algorithmically than by hand-coding heuristics.

Third, we examined the cases in which the algorithm should use the same action more than once when creating agent copies, and cases in which it should use a diverse action set.

This formalism, while more restrictive than the general multiagent case, seems likely to be useful in Mobile Wireless Networks and other domains as well.

\begin{acks}
This material is based upon work supported by the National Science Foundation under Grant No. 2154190. Any opinions, findings, and conclusions or recommendations expressed in this material are those of the author(s) and do not necessarily reflect the views of the National Science Foundation.
\end{acks}

\bibliographystyle{ACM-Reference-Format} 
\bibliography{bibs/Multicopy,bibs/networks,bibs/ReviewerReferencesALA}


\begin{thebibliography}{25}


\ifx \showCODEN    \undefined \def \showCODEN     #1{\unskip}     \fi
\ifx \showDOI      \undefined \def \showDOI       #1{#1}\fi
\ifx \showISBNx    \undefined \def \showISBNx     #1{\unskip}     \fi
\ifx \showISBNxiii \undefined \def \showISBNxiii  #1{\unskip}     \fi
\ifx \showISSN     \undefined \def \showISSN      #1{\unskip}     \fi
\ifx \showLCCN     \undefined \def \showLCCN      #1{\unskip}     \fi
\ifx \shownote     \undefined \def \shownote      #1{#1}          \fi
\ifx \showarticletitle \undefined \def \showarticletitle #1{#1}   \fi
\ifx \showURL      \undefined \def \showURL       {\relax}        \fi
\providecommand\bibfield[2]{#2}
\providecommand\bibinfo[2]{#2}
\providecommand\natexlab[1]{#1}
\providecommand\showeprint[2][]{arXiv:#2}

\bibitem[\protect\citeauthoryear{Albrecht, Christianos, and Sch\"afer}{Albrecht
  et~al\mbox{.}}{2024}]%
        {marl-book}
\bibfield{author}{\bibinfo{person}{Stefano~V. Albrecht},
  \bibinfo{person}{Filippos Christianos}, {and} \bibinfo{person}{Lukas
  Sch\"afer}.} \bibinfo{year}{2024}\natexlab{}.
\newblock \bibinfo{booktitle}{\emph{Multi-Agent Reinforcement Learning:
  Foundations and Modern Approaches}}.
\newblock \bibinfo{publisher}{MIT Press}.
\newblock
\urldef\tempurl%
\url{https://www.marl-book.com}
\showURL{%
\tempurl}


\bibitem[\protect\citeauthoryear{Arulkumaran, Deisenroth, Brundage, and
  Bharath}{Arulkumaran et~al\mbox{.}}{2017}]%
        {deeprl}
\bibfield{author}{\bibinfo{person}{Kai Arulkumaran},
  \bibinfo{person}{Marc~Peter Deisenroth}, \bibinfo{person}{Miles Brundage},
  {and} \bibinfo{person}{Anil~Anthony Bharath}.}
  \bibinfo{year}{2017}\natexlab{}.
\newblock \showarticletitle{Deep Reinforcement Learning: A Brief Survey}.
\newblock \bibinfo{journal}{\emph{IEEE Signal Processing Magazine}}
  \bibinfo{volume}{34}, \bibinfo{number}{6} (\bibinfo{year}{2017}),
  \bibinfo{pages}{26--38}.
\newblock
\urldef\tempurl%
\url{https://doi.org/10.1109/MSP.2017.2743240}
\showDOI{\tempurl}


\bibitem[\protect\citeauthoryear{Bellemare, Dabney, and Rowland}{Bellemare
  et~al\mbox{.}}{[n.d.]}]%
        {bellemare_distributional_2023}
\bibfield{author}{\bibinfo{person}{Marc~G. Bellemare}, \bibinfo{person}{Will
  Dabney}, {and} \bibinfo{person}{Mark Rowland}.}
  \bibinfo{year}{[n.d.]}\natexlab{}.
\newblock \bibinfo{booktitle}{\emph{Distributional Reinforcement Learning}}.
\newblock \bibinfo{publisher}{{MIT} Press}.
\newblock
\urldef\tempurl%
\url{https://www.distributional-rl.org/}
\showURL{%
\tempurl}


\bibitem[\protect\citeauthoryear{DeNero and Klein}{DeNero and Klein}{2010}]%
        {pacman_2010}
\bibfield{author}{\bibinfo{person}{John DeNero} {and} \bibinfo{person}{Dan
  Klein}.} \bibinfo{year}{2010}\natexlab{}.
\newblock \showarticletitle{Teaching Introductory Artificial Intelligence with
  Pac-Man}.
\newblock \bibinfo{journal}{\emph{Proceedings of the AAAI Conference on
  Artificial Intelligence}} \bibinfo{volume}{24}, \bibinfo{number}{3}
  (\bibinfo{date}{Jul.} \bibinfo{year}{2010}), \bibinfo{pages}{1885--1889}.
\newblock
\urldef\tempurl%
\url{https://doi.org/10.1609/aaai.v24i3.18829}
\showDOI{\tempurl}


\bibitem[\protect\citeauthoryear{Fulda and Ventura}{Fulda and Ventura}{2007}]%
        {fulda_predicting_2007}
\bibfield{author}{\bibinfo{person}{Nancy Fulda} {and} \bibinfo{person}{Dan
  Ventura}.} \bibinfo{year}{2007}\natexlab{}.
\newblock \showarticletitle{Predicting and preventing coordination problems in
  cooperative Q-learning systems}. In \bibinfo{booktitle}{\emph{Proceedings of
  the 20th international joint conference on Artifical intelligence}} (San
  Francisco, {CA}, {USA}, 2007-01-06) \emph{(\bibinfo{series}{{IJCAI}'07})}.
  \bibinfo{publisher}{Morgan Kaufmann Publishers Inc.},
  \bibinfo{pages}{780--785}.
\newblock


\bibitem[\protect\citeauthoryear{Gronauer and Diepold}{Gronauer and
  Diepold}{2022}]%
        {gronauer_multi-agent_2022}
\bibfield{author}{\bibinfo{person}{Sven Gronauer} {and} \bibinfo{person}{Klaus
  Diepold}.} \bibinfo{year}{2022}\natexlab{}.
\newblock \showarticletitle{Multi-agent deep reinforcement learning: a survey}.
\newblock \bibinfo{journal}{\emph{Artificial Intelligence Review}}
  \bibinfo{volume}{55}, \bibinfo{number}{2} (\bibinfo{year}{2022}),
  \bibinfo{pages}{895--943}.
\newblock
\showISSN{1573-7462}
\urldef\tempurl%
\url{https://doi.org/10.1007/s10462-021-09996-w}
\showDOI{\tempurl}


\bibitem[\protect\citeauthoryear{Haarnoja, Tang, Abbeel, and Levine}{Haarnoja
  et~al\mbox{.}}{[n.d.]}]%
        {haarnoja_reinforcement_2017}
\bibfield{author}{\bibinfo{person}{Tuomas Haarnoja}, \bibinfo{person}{Haoran
  Tang}, \bibinfo{person}{Pieter Abbeel}, {and} \bibinfo{person}{Sergey
  Levine}.} \bibinfo{year}{[n.d.]}\natexlab{}.
\newblock \bibinfo{title}{Reinforcement Learning with Deep Energy-Based
  Policies}.
\newblock
\newblock
\urldef\tempurl%
\url{https://doi.org/10.48550/arXiv.1702.08165}
\showDOI{\tempurl}
\showeprint[arxiv]{1702.08165 [cs]}


\bibitem[\protect\citeauthoryear{Hayes}{Hayes}{[n.d.]}]%
        {hayes_brief_2023}
\bibfield{author}{\bibinfo{person}{Conor~F Hayes}.}
  \bibinfo{year}{[n.d.]}\natexlab{}.
\newblock \showarticletitle{A Brief Guide to Multi-Objective Reinforcement
  Learning and Planning}.
\newblock  (\bibinfo{year}{[n.\,d.]}).
\newblock


\bibitem[\protect\citeauthoryear{Hayes, Howley, and Mannion}{Hayes
  et~al\mbox{.}}{[n.d.]}]%
        {hayes_dynamic_nodate}
\bibfield{author}{\bibinfo{person}{Conor~F Hayes}, \bibinfo{person}{Enda
  Howley}, {and} \bibinfo{person}{Patrick Mannion}.}
  \bibinfo{year}{[n.d.]}\natexlab{}.
\newblock \showarticletitle{Dynamic Thresholded Lexicographic Ordering}.
\newblock  (\bibinfo{year}{[n.\,d.]}).
\newblock


\bibitem[\protect\citeauthoryear{Jain, Fall, and Patra}{Jain
  et~al\mbox{.}}{2004}]%
        {Jain2004:DTN}
\bibfield{author}{\bibinfo{person}{S. Jain}, \bibinfo{person}{K. Fall}, {and}
  \bibinfo{person}{R. Patra}.} \bibinfo{year}{2004}\natexlab{}.
\newblock \showarticletitle{Routing in a delay tolerant network}. In
  \bibinfo{booktitle}{\emph{Proc. of SIGCOMM}}.
\newblock


\bibitem[\protect\citeauthoryear{Khorasgani, Wang, Tang, and Gupta}{Khorasgani
  et~al\mbox{.}}{2021}]%
        {khorasgani_k-nearest_2022}
\bibfield{author}{\bibinfo{person}{H. Khorasgani}, \bibinfo{person}{H. Wang},
  \bibinfo{person}{H. Tang}, {and} \bibinfo{person}{C. Gupta}.}
  \bibinfo{year}{2021}\natexlab{}.
\newblock \showarticletitle{K-nearest Multi-agent Deep Reinforcement Learning
  for Collaborative Tasks with a Variable Number of Agents}. In
  \bibinfo{booktitle}{\emph{2021 IEEE International Conference on Big Data (Big
  Data)}}. \bibinfo{publisher}{IEEE Computer Society}, \bibinfo{address}{Los
  Alamitos, CA, USA}, \bibinfo{pages}{3883--3889}.
\newblock
\urldef\tempurl%
\url{https://doi.org/10.1109/BigData52589.2021.9671691}
\showDOI{\tempurl}


\bibitem[\protect\citeauthoryear{Manfredi, Wolfe, Zhang, and Wang}{Manfredi
  et~al\mbox{.}}{[n.d.]}]%
        {manfredi_learning_2024}
\bibfield{author}{\bibinfo{person}{Victoria Manfredi},
  \bibinfo{person}{Alicia~P. Wolfe}, \bibinfo{person}{Xiaolan Zhang}, {and}
  \bibinfo{person}{Bing Wang}.} \bibinfo{year}{[n.d.]}\natexlab{}.
\newblock \showarticletitle{Learning an adaptive forwarding strategy for mobile
  wireless networks: resource usage vs. latency}.
\newblock  (\bibinfo{year}{[n.\,d.]}).
\newblock
\showISSN{1573-0565}
\urldef\tempurl%
\url{https://doi.org/10.1007/s10994-024-06601-3}
\showDOI{\tempurl}


\bibitem[\protect\citeauthoryear{Moffaert and Nowé}{Moffaert and
  Nowé}{[n.d.]}]%
        {moffaert_multi-objective_2014}
\bibfield{author}{\bibinfo{person}{Kristof~Van Moffaert} {and}
  \bibinfo{person}{Ann Nowé}.} \bibinfo{year}{[n.d.]}\natexlab{}.
\newblock \showarticletitle{Multi-Objective Reinforcement Learning using Sets
  of Pareto Dominating Policies}.
\newblock  \bibinfo{volume}{15}, \bibinfo{number}{107}
  (\bibinfo{year}{[n.\,d.]}), \bibinfo{pages}{3663--3692}.
\newblock
\showISSN{1533-7928}
\urldef\tempurl%
\url{http://jmlr.org/papers/v15/vanmoffaert14a.html}
\showURL{%
\tempurl}


\bibitem[\protect\citeauthoryear{Mossalam, Assael, Roijers, and
  Whiteson}{Mossalam et~al\mbox{.}}{[n.d.]}]%
        {mossalam_multi-objective_2016}
\bibfield{author}{\bibinfo{person}{Hossam Mossalam}, \bibinfo{person}{Yannis~M.
  Assael}, \bibinfo{person}{Diederik~M. Roijers}, {and} \bibinfo{person}{Shimon
  Whiteson}.} \bibinfo{year}{[n.d.]}\natexlab{}.
\newblock \bibinfo{title}{Multi-Objective Deep Reinforcement Learning}.
\newblock
\newblock
\urldef\tempurl%
\url{https://doi.org/10.48550/arXiv.1610.02707}
\showDOI{\tempurl}
\showeprint[arxiv]{1610.02707 [cs]}


\bibitem[\protect\citeauthoryear{Nguyen, Nguyen, Vamplew, Nahavandi, Dazeley,
  and Lim}{Nguyen et~al\mbox{.}}{[n.d.]}]%
        {nguyen_multi-objective_2020}
\bibfield{author}{\bibinfo{person}{Thanh~Thi Nguyen}, \bibinfo{person}{Ngoc~Duy
  Nguyen}, \bibinfo{person}{Peter Vamplew}, \bibinfo{person}{Saeid Nahavandi},
  \bibinfo{person}{Richard Dazeley}, {and} \bibinfo{person}{Chee~Peng Lim}.}
  \bibinfo{year}{[n.d.]}\natexlab{}.
\newblock \showarticletitle{A multi-objective deep reinforcement learning
  framework}.
\newblock   \bibinfo{volume}{96} (\bibinfo{year}{[n.\,d.]}),
  \bibinfo{pages}{103915}.
\newblock
\showISSN{0952-1976}
\urldef\tempurl%
\url{https://doi.org/10.1016/j.engappai.2020.103915}
\showDOI{\tempurl}


\bibitem[\protect\citeauthoryear{Oroojlooy and Hajinezhad}{Oroojlooy and
  Hajinezhad}{2023}]%
        {oroojlooy2023review}
\bibfield{author}{\bibinfo{person}{Afshin Oroojlooy} {and}
  \bibinfo{person}{Davood Hajinezhad}.} \bibinfo{year}{2023}\natexlab{}.
\newblock \showarticletitle{A review of cooperative multi-agent deep
  reinforcement learning}.
\newblock \bibinfo{journal}{\emph{Applied Intelligence}} \bibinfo{volume}{53},
  \bibinfo{number}{11} (\bibinfo{year}{2023}), \bibinfo{pages}{13677--13722}.
\newblock


\bibitem[\protect\citeauthoryear{Rummery and Niranjan}{Rummery and
  Niranjan}{1994}]%
        {sarsa_rummery1994}
\bibfield{author}{\bibinfo{person}{G.~A. Rummery} {and} \bibinfo{person}{M.
  Niranjan}.} \bibinfo{year}{1994}\natexlab{}.
\newblock \bibinfo{booktitle}{\emph{On-line Q-learning using connectionist
  systems.}}
\newblock \bibinfo{type}{{T}echnical {R}eport} CUED/F-INFENG/TR 166.
  \bibinfo{institution}{Engineering Department, Cambridge University}.
\newblock


\bibitem[\protect\citeauthoryear{Schroeder~de Witt, Foerster, Farquhar, Torr,
  Boehmer, and Whiteson}{Schroeder~de Witt et~al\mbox{.}}{2019}]%
        {schroeder_de_witt_multi-agent_2019}
\bibfield{author}{\bibinfo{person}{Christian Schroeder~de Witt},
  \bibinfo{person}{Jakob Foerster}, \bibinfo{person}{Gregory Farquhar},
  \bibinfo{person}{Philip Torr}, \bibinfo{person}{Wendelin Boehmer}, {and}
  \bibinfo{person}{Shimon Whiteson}.} \bibinfo{year}{2019}\natexlab{}.
\newblock \showarticletitle{Multi-Agent Common Knowledge Reinforcement
  Learning}. In \bibinfo{booktitle}{\emph{Advances in Neural Information
  Processing Systems}}, Vol.~\bibinfo{volume}{32}. \bibinfo{publisher}{Curran
  Associates, Inc.}
\newblock
\urldef\tempurl%
\url{https://proceedings.neurips.cc/paper_files/paper/2019/hash/f968fdc88852a4a3a27a81fe3f57bfc5-Abstract.html}
\showURL{%
\tempurl}


\bibitem[\protect\citeauthoryear{Spyropoulos, Psounis, and
  Raghavendra}{Spyropoulos et~al\mbox{.}}{2008}]%
        {multicopy-spyro}
\bibfield{author}{\bibinfo{person}{Thrasyvoulos Spyropoulos},
  \bibinfo{person}{Konstantinos Psounis}, {and} \bibinfo{person}{Cauligi~S
  Raghavendra}.} \bibinfo{year}{2008}\natexlab{}.
\newblock \showarticletitle{Efficient routing in intermittently connected
  mobile networks: The multiple-copy case}.
\newblock \bibinfo{journal}{\emph{IEEE/ACM transactions on networking}}
  \bibinfo{volume}{16}, \bibinfo{number}{1} (\bibinfo{year}{2008}),
  \bibinfo{pages}{77--90}.
\newblock


\bibitem[\protect\citeauthoryear{Sutton and Barto}{Sutton and Barto}{2018}]%
        {sutton_reinforcement_2018}
\bibfield{author}{\bibinfo{person}{Richard~S Sutton} {and}
  \bibinfo{person}{Andrew~G Barto}.} \bibinfo{year}{2018}\natexlab{}.
\newblock \bibinfo{booktitle}{\emph{Reinforcement Learning: An Introduction}
  (\bibinfo{edition}{2nd} ed.)}.
\newblock \bibinfo{publisher}{{MIT} Press}.
\newblock
\urldef\tempurl%
\url{http://www.incompleteideas.net/book/the-book-2nd.html}
\showURL{%
\tempurl}


\bibitem[\protect\citeauthoryear{Tercan and Prabhu}{Tercan and
  Prabhu}{[n.d.]}]%
        {tercan_thresholded_2022}
\bibfield{author}{\bibinfo{person}{Alperen Tercan} {and}
  \bibinfo{person}{Vinayak Prabhu}.} \bibinfo{year}{[n.d.]}\natexlab{}.
\newblock \showarticletitle{Thresholded Lexicographic Ordered Multi-Objective
  Reinforcement Learning}.
\newblock  (\bibinfo{year}{[n.\,d.]}).
\newblock
\urldef\tempurl%
\url{https://openreview.net/forum?id=mmFtinp4wQ_}
\showURL{%
\tempurl}


\bibitem[\protect\citeauthoryear{Tie, Venkataramani, and Balasubramanian}{Tie
  et~al\mbox{.}}{2011}]%
        {tie2011r3}
\bibfield{author}{\bibinfo{person}{Xiaozheng Tie}, \bibinfo{person}{Arun
  Venkataramani}, {and} \bibinfo{person}{Aruna Balasubramanian}.}
  \bibinfo{year}{2011}\natexlab{}.
\newblock \showarticletitle{R3: Robust replication routing in wireless networks
  with diverse connectivity characteristics}. In
  \bibinfo{booktitle}{\emph{Proceedings of the 17th annual international
  conference on Mobile computing and networking}}. \bibinfo{pages}{181--192}.
\newblock


\bibitem[\protect\citeauthoryear{Vieillard, Pietquin, and Geist}{Vieillard
  et~al\mbox{.}}{[n.d.]}]%
        {vieillard_munchausen_2020}
\bibfield{author}{\bibinfo{person}{Nino Vieillard}, \bibinfo{person}{Olivier
  Pietquin}, {and} \bibinfo{person}{Matthieu Geist}.}
  \bibinfo{year}{[n.d.]}\natexlab{}.
\newblock \showarticletitle{Munchausen Reinforcement Learning}. In
  \bibinfo{booktitle}{\emph{Advances in Neural Information Processing Systems}}
  (2020), Vol.~\bibinfo{volume}{33}. \bibinfo{publisher}{Curran Associates,
  Inc.}, \bibinfo{pages}{4235--4246}.
\newblock
\urldef\tempurl%
\url{https://proceedings.neurips.cc/paper/2020/hash/2c6a0bae0f071cbbf0bb3d5b11d90a82-Abstract.html}
\showURL{%
\tempurl}


\bibitem[\protect\citeauthoryear{Watkins and Dayan}{Watkins and Dayan}{1992}]%
        {watkins1992}
\bibfield{author}{\bibinfo{person}{C.~J. C.~H. Watkins} {and}
  \bibinfo{person}{P. Dayan}.} \bibinfo{year}{1992}\natexlab{}.
\newblock \showarticletitle{Q-learning}.
\newblock \bibinfo{journal}{\emph{Machine Learning}} \bibinfo{volume}{8},
  \bibinfo{number}{3-4} (\bibinfo{year}{1992}), \bibinfo{pages}{279--292}.
\newblock


\bibitem[\protect\citeauthoryear{Wong, Bäck, Kononova, and Plaat}{Wong
  et~al\mbox{.}}{2023}]%
        {wong_deep_2023}
\bibfield{author}{\bibinfo{person}{Annie Wong}, \bibinfo{person}{Thomas Bäck},
  \bibinfo{person}{Anna~V. Kononova}, {and} \bibinfo{person}{Aske Plaat}.}
  \bibinfo{year}{2023}\natexlab{}.
\newblock \showarticletitle{Deep multiagent reinforcement learning: challenges
  and directions}.
\newblock \bibinfo{journal}{\emph{Artificial Intelligence Review}}
  \bibinfo{volume}{56}, \bibinfo{number}{6} (\bibinfo{year}{2023}),
  \bibinfo{pages}{5023--5056}.
\newblock
\showISSN{1573-7462}
\urldef\tempurl%
\url{https://doi.org/10.1007/s10462-022-10299-x}
\showDOI{\tempurl}


\end{thebibliography}

   \end{document}